\documentclass[final]{IEEEtran}
\usepackage{amsthm,amssymb,graphicx,multirow,amsmath,color,amsfonts}%,ulem}
\usepackage[caption=false,font=footnotesize]{subfig}
\usepackage[update,prepend]{epstopdf}
\usepackage[noadjust]{cite}
\usepackage[latin1]{inputenc}
\usepackage{tikz}
\usepackage{bbm} % for \mathbbm{1}
\usepackage{pdfpages}
\usepackage{multirow}
\usepackage{comment}
\usepackage[ruled,vlined]{algorithm2e}
\usepackage{url}
\makeatother

\def\nb0{{\mathbf{0}}}
\def\nb1{{\mathbf{1}}}

\def\ncalH{{\mathcal{H}}}

\def\ncalN{{\mathcal{N}}}
\def\ncalO{{\mathcal{O}}}
\def\ncalP{{\mathcal{P}}}

\def\ncalS{{\mathcal{S}}}

\def\ncalW{{\mathcal{W}}}

\def\ncalZ{{\mathcal{Z}}}

\def\nbbN{{\mathbb{N}}}

\def\nbbR{{\mathbb{R}}}

\newtheorem{definition}{Definition}

\newtheorem{theorem}{Theorem}

\newtheorem{remark}{Remark}

\def\E{\mathbb{E}}

\allowdisplaybreaks

\usepackage{setspace}

\begin{document}

\title{Online Meta-Learning for Scene-Diverse Waveform-Agile Radar Target Tracking
{\footnotesize \textsuperscript{}}
\thanks{$^*$C.E. Thornton and R.M. Buehrer are with Wireless@VT, Bradley Department of ECE, Virginia Tech, Blacksburg, VA, 24061. (\textit{Emails: \{thorntonc, buehrer\}@vt.edu}). $^\ddagger$A.F. Martone is with the U.S Army Research Laboratory, Adelphi, MD 20783. (\textit{Email: anthony.f.martone.civ@mail.mil}). The support of the U.S Army Research Office (ARO) is gratefully acknowledged.}}
\author{\IEEEauthorblockN{Charles E. Thornton$^{*}$, R. Michael Buehrer$^{*}$, and Anthony F. Martone$^{\ddagger}$}}

\maketitle

\begin{abstract}	
A fundamental problem for waveform-agile radar systems is that the true environment is unknown, and transmission policies which perform well for a particular tracking instance may be sub-optimal for another. Additionally, there is a limited time window for each target track, and the radar must learn an effective strategy from a sequence of measurements in a timely manner. This paper studies a Bayesian meta-learning model for radar waveform selection which seeks to learn an inductive bias to quickly optimize tracking performance across a class of radar scenes. We cast the waveform selection problem in the framework of sequential Bayesian inference, and introduce a contextual bandit variant of the recently proposed meta-Thompson Sampling algorithm, which learns an inductive bias in the form of a prior distribution. Each track is treated as an instance of a contextual bandit learning problem, coming from a \emph{task distribution}. We show that the meta-learning process results in an appreciably faster learning, resulting in significantly fewer lost tracks than a conventional learning approach equipped with an uninformative prior.

%Sequential Bayesian learning has been identified as a key component of emerging cognitive radar systems, especially those which engage in target tracking. Since target measurements are observed sequentially, and decisions regarding the transmission waveform must be made on a timely basis, sample efficient and computationally feasible online learning algorithms are of significant interest for practical implementations. This work develops a meta-learning framework for online waveform selection in target tracking systems, with the goal of learning \emph{generalizable} strategies across a broad range of sensing environments by learning an inductive bias in the form of a prior distribution. In this formulation, each track is treated as an instance of a contextual bandit learning problem, coming from a \emph{task distribution}. Over time, the proposed meta-Thompson Sampling algorithm is shown to learn a prior distribution over candidate waveforms that results in favorable performance when compared to Thompson Sampling used with a randomly initialized or uninformative prior. We show that the meta-learning process results in a significantly reduced number of lost tracks compared to conventional learning.
\end{abstract}

\vspace{-.4cm}
\section{Introduction}
A cognitive radar is a closed-loop, adaptive, information gathering system. The cognitive radar reduces uncertainty about its environment by sequentially probing the surroundings and applying a mix of signal and information processing techniques to improve performance over time. The goal of such a system is to employ a transmission and reception strategy that is, in some sense, optimal for the radar's sensing environment. By nature, some parts of the sensing environment, or \emph{scene}, will be unknown to the radar. In order to make decisions regarding signal transmission and processing, a model of the underlying environment's dynamics is required. This model can either be pre-conceived, or learned directly from received data. The ability to learn is certainly favorable, as fewer assumptions regarding the environment are necessary. Learning has also been identified by Haykin as a key component of cognitive radar \cite{Haykin2012}, and has been explored in the literature by several authors. However, a major concern surrounding data-driven learning is that the radar must start with limited prior knowledge, and performance may be poor until sufficient experience has been accumulated. 

For critical applications such as target tracking, it is important to maintain sufficient measurement quality throughout the duration of the sensing objective. Additionally, tracking radars often engage in an ongoing sequence of tracks, where the duration of each individual tracking interval is limited by the amount of time the object is within the radar's sensing range. Thus, cognitive tracking radars are expected to quickly learn strategies which generalize across a variety of sensing environments.

Meta-learning is a practical machine learning approach to ensure that a decision-making agent is both flexible and data efficient \cite{Ortega2019}. In general, the goal of meta-learning is to acquire an \emph{inductive bias} by exposing the agent to a distribution of related tasks. This is in contrast to a conventional machine-learning approach, where the agent is exposed to a single task, and is concerned with optimizing performance for that particular scenario. Thus, a key assumption for meta-learning is that the agent is embedded in an underlying environment, from which related tasks are generated. For tracking radar, this assumption is reasonable, as the scene will often vary from track to track, but can be expected to retain certain fundamental properties. 

Meta-learning has recently shown promise for a multitude of wireless applications, such as demodulation from limited pilot symbols \cite{Park2021} and power control over distributed communication networks \cite{Nikoloska2021}. However, we argue that due to the ease with which radar feedback can be obtained, as well as the sequential nature of the tracking problem, cognitive radar is particularly well-posed to benefit from meta-learning techniques.

\emph{Contributions:} We develop an online meta-learning model for waveform-agile target tracking. Our approach follows the Bayesian interpretation of meta-learning introduced by Baxter \cite{Baxter1998}. We build on the meta-Thompson Sampling algorithm introduced in \cite{Kveton2021}, by introducing a linear contextual bandit model of meta-Thompson Sampling applicable to the cognitive radar problem. We demonstrate that this approach is effective for target tracking, and results in a significantly reduced number of lost tracks when compared to a traditional learning approach with an uninformative prior.

\section{Waveform-Agile Tracking Problem}
In this section, we describe the problem of waveform-agile tracking with a cognitive radar system. We begin by describing a finite-state target channel model for a radar environment with memory. We then proceed to a description of how the waveform selection problem can be cast in the form of Bayesian inference and solved using the well-studied linear contextual bandit formulation. We finally describe the sensitivity of Bayesian learning algorithms to the choice of prior, in order to show why meta-learning can provide substantial performance improvements in tracking problems.
\label{se:problem}
\subsection{Finite State Target Channel Model}
Each discrete time index\footnote{For convenience, we consider discrete time with finite observation and waveform sets. The extension to continuous domains comes with a loss of generality, and requires more sophisticated assumptions.} $(k)_{k=1}^{n} \in \mathbb{N}_{+}$, the radar scene is said to be in state $s_{k} $, which\footnote{We use the notation $s_{1}^{k}$ to denote the sequence of states from time $1$ to $k$.} takes values in finite alphabet $\mathcal{S}$. The state describes relative losses due to the scattering effects of the propagation environment, as well as from the target. The state temporally evolves according to a discrete stochastic process $\{s_{k}\}_{k \in \mathbb{N}}$, with memory. The state generating process has transition probabilities $P(s_{k+1}|s_{1}^{k})$ which are unknown to the radar \emph{a priori}. We will refer to a particular state transition model as the \emph{task}, to remain consistent with the meta-learning literature. In this formulation, we assume the scene's state transitions occur independent of the radar's transmissions.

By nature, the radar is a measurement system, and cannot observe the true state directly. The radar instead makes an observation $o_{k} \in \mathcal{O}$ at each time step. We assume $\mathcal{O} = \mathcal{S}$, which is reasonable since the radar should have an idea of which hypotheses to expect regarding channel conditions. The observation process is governed by the probability kernel $P(o_{k}|s_{k})$. After receiving $o_{k}$, The radar must select a waveform $w_{k}$ from a finite alphabet of waveforms $\mathcal{W}$. The radar wishes to measure a random vector of target parameters $z_{k} \in \mathcal{Z}$. We have now established the tools to define our finite state target channel (FSTC), which the radar will interact with during a target track.

\begin{definition}[FSTC]
	The FSTC is given by the tuple $(\ncalW \times \ncalS^{k}, P(z|w,s_{1}^{k}), \ncalZ)$, where $\ncalS$ is a finite alphabet of states, $\ncalW$ is a finite waveform alphabet, $\ncalZ$ is a finite set of possible target parameter vectors, and $P: \ncalW \times \ncalS \mapsto \ncalZ$ is a stochastic matrix which maps channel inputs to target parameter measurements.
\end{definition}

\begin{remark}
	Our channel model is a generalization of a FSTC proposed by Bell \cite{Bell1988}, as the current state generating process is not assumed to be Markov, and may depend on the entire past state sequence.
\end{remark}

\begin{remark}
	A conventional learning approach would seek to optimize the choice of waveforms for a fixed, but a priori unknown, FSTC. Meta-learning aims to find an inductive bias which allows for efficient learning across a \textbf{class} of FSTCs drawn from a common distribution.
\end{remark}

Based on the selected waveform and underlying state, the radar observes a loss\footnote{We describe our specific choice of loss function in Section \ref{se:numerical}.} $\ell_{k}: \ncalW \times \ncalS \mapsto \mathbb{R}$. We assume that the radar uses a fixed processing scheme, and thus the target parameter estimate $z_{k}$ will depend only on the pair $(w_{k},s_{k})$. 

The radar is tracking the target parameter values over a finite time horizon $n$. The goal of the learning problem is to select a sequence of waveforms $\{w_{k}\}_{k=1}^{n}$ which minimize $\E[\ell_{k}(w_{k},s_{k})]$, where the expectation is over the perceived randomness in the state generating process. Unfortunately, this expectation is not computable as both the loss mapping $\ell_{k}$ and the process $\{s_{k}\}$ are unknown. Instead, the radar aims to minimize the empirical average loss
\begin{equation*}
	J_{n} = \frac{1}{n} \sum_{k=1}^{n} \ell_{k}(w_{k},s_{k}),
\end{equation*}
which is challenging since each $s_{k}$ is only partially observed through $o_{k}$, and the structure of $\ell_{k}$ is unknown to the radar \emph{a priori}. The radar must select waveforms using the history of observations, selected waveforms, and received losses up to step $k-1$ given by
\begin{equation*}
	\ncalH_{k-1} = \{(o_{i},w_{i},\ell_{i})\}_{i=1}^{k-1}
\end{equation*}

Thus, the radar must gather enough information about the task to select optimal waveforms, while minimizing the total number of sub-optimal waveforms selected. This is the classic dilemma of \emph{exploration} and \emph{exploitation}, which has been studied in the general sense \cite{Lattimore2020}, as well as in the context of radar waveform selection \cite{Thornton2021}. We describe the sequential learning problem in more detail in the next section.

\subsection{Bayesian Learning Problem}
A common view of the sequential decision problem is the Bayesian interpretation, which makes decisions with respect to the posterior probability of being the optimal action (see Section 2.2 of \cite{Baxter1998}). This approach assumes the task can be described by a set of probability distributions $\{P_{\theta}\}$ parameterized by $\theta \in \Theta$, where $\Theta$ is a subset of $\nbbR^{d}$. An example of such a case is the well-known \emph{stochastic linear bandit} problem (Chapter 19 of \cite{Lattimore2020}), where we define $\varphi: \ncalO \times \ncalW \mapsto \nbbR^{d}$ to be a feature mapping\footnote{We describe our specific choice of feature mappings in Section \ref{se:numerical}.}, describing aspects of the physical scene, and the relation
\begin{equation*}
	\ell_{k} = \langle \theta^{*}, \varphi(o_{i},w_{i}) \rangle + \eta_{k},
\end{equation*}
holds for all $(o_{i},w_{i}) \in \ncalO \times \ncalW$, where $\eta_{k}$ is a sub-Gaussian random disturbance. In this case, the vector $\theta$ is sufficient to predict $\E[\ell_{k}]$ for all $(o_{k},w_{k})$, and the goal is to use data $\ncalH$ to learn a posterior distribution over the parameter $P(\theta|\ncalH)$. This is the base learning problem we will study in the remainder of this paper.

The radar starts with a prior over possible parameter values $P(\theta)$. Using the observed data $\ncalH$, the radar then updates its prior distribution to a posterior distribution by applying Bayes rule
\begin{align*}
	P(\theta|\mathcal{H}) &= \frac{P(\mathcal{H}|\theta)P(\theta)}{P(\mathcal{H})} \\
						  &= \frac{\prod_{i=1}^{n}P(\{w_{i},o_{i},\ell_{i}\}|\theta)P(\theta)}{\int_{\Theta}P(\mathcal{H}|\theta)P(\theta)d\theta}.
\end{align*}

Since the radar is interested in predicting the loss given observations, we can factor $P(o,w,\ell|\theta)$ into $P(o,w)P(\ell|o,w;\theta)$. The posterior $P(\theta|\ncalH)$ can then be used to predict the loss $\ell_{k}$ associated with a waveform, observation pair $(w_{k},o_{k})$ by taking an average
\begin{equation*}
	P(\ell_{k}|(w_{k},o_{k});\ncalH) = \int_{\Theta} P(\ell_{k}|(w_{k},o_{k});\theta) P(\theta|\ncalH) d\theta.
\end{equation*}

As $k$ increases, predictions made using the history $\ncalH_{k-1}$ are expected to improve. One way of interpreting this is through the KL divergence between the true distribution $P_{\theta^{*}}$ and the posterior estimated using $\ncalH_{n}$, $P_{n}$, given by
\begin{align*}
 &D_{\texttt{KL}}(P_{\theta^{*}}||P_{n})\\ &= \int_{\ncalH} P(w,o,\ell|\theta^{*}) \log \left( \frac{P(w,o,\ell|\theta^{*})}{P(w,o,\ell|\ncalH)} \right) dw \; do \; d\ell\\
										   &= \int_{\ncalH} P(w,o)P(\ell|w,o;\theta^{*}) \log \left( \frac{P(\ncalH|w,o;\theta^{*})}{P(\ell|w,o;\ncalH)} \right) dw \; do \; d\ell.
\end{align*}

In the case of a long time-horizon, $n$, it has been shown (\cite{Clarke1990}) that under certain restrictions, if $\Theta$ is a compact subset of $\mathbb{R}^{d}$,
\begin{equation*}
	D_{\texttt{KL}}(P_{\theta^{*}}||P_{n}) = \frac{d}{n}+ o \left( \frac{1}{n} \right),
\end{equation*}
assuming sufficient exploration of the space $\ncalW \times \ncalS$. Thus, we see that as long as the waveform selection problem can be parameterized by $\theta \in \Theta$ and is relatively well-behaved, we can come very close to the best possible performance given sufficient experience captured in $\ncalH$. However, a significant challenge is selecting the prior distribution $P(\theta)$ such that a limited amount of data is needed to learn the true distribution. This is especially true for problems over a limited time horizon, such as target tracking. Ideally, we would seek to minimize $D_{\texttt{KL}}(P_{\theta^{*}}||P(\theta))$. However, since the optimal parameter $\theta^{*}$ is by necessity unknown, the prior is usually selected to be an uninformative or randomly initialized distribution. While selecting actions according to their posterior probability of being optimal will still result in good performance in the asymptotic regime, this may result in highly suboptimal near-term behavior.

\begin{definition}[Thompson Sampling]
	Thompson sampling involves selecting the waveform $w^{*} \in \ncalW$ such that 
	\begin{equation}
		\label{eq:tspolicy}
		\int_{\Theta} \mathbbm{1} \left[ \E[\ell|w^{*},o,\theta] = \min_{w'} \E[\ell|w',o,\theta] \right] P(\theta|\ncalH) d\theta,
	\end{equation}
where $\mathbbm{1}[\cdot]$ is the indicator function. Details of a practical algorithm to achieve this policy in the linear contextual bandit setting can be found in \cite{Agrawal2013}. Details of a context-aware waveform selection strategy based on Thompson Sampling can be found in \cite{Thornton2021}.
\end{definition}

\begin{definition}[Bayesian Regret]
	The expected, or Bayesian, $n$-round regret of a decision strategy is given by
	\begin{equation*}
		\operatorname{BR}^{*}_{n} \triangleq \E \left[\sum_{k=1}^{n} \ell_{k} - \sum_{k=1}^{n} \min_{w^{*} \in \mathcal{W}} \langle \theta, w^{*} \rangle \right],
	\end{equation*}
	which is an indication of the average performance of a decision algorithm. However, an algorithm with modest Bayesian regret may perform poorly in particular scenarios.
\end{definition}

\begin{theorem}[Prior Dependent Upper/Lower Bound for Bayesian Bandit \cite{Lattimore2020}]
	For any prior $Q$, the Bayesian regret of a $k$-armed bandit satisfies
	\begin{equation*}
		\operatorname{BR}^{*}_{n}(Q) \leq C \sqrt{kn},
	\end{equation*}
	where $C > 0$ is a universal constant. Further, there exists a prior $Q$ such that
	\begin{equation*}
		\operatorname{BR}^{*}_{n}(Q) \geq c \sqrt{kn},
	\end{equation*}
	where $c > 0$ is a universal constant. Thus, we see that $\sup_{Q} \operatorname{BR}^{*}_{n}(Q) = \Theta(\sqrt{kn})$, where $f(x) = \Theta(g(x))$ denotes that $f(x)$ is bounded both above and below by $g(x)$ asymptotically.
\end{theorem}

Thus, we see that for the case of small time-horizon $n$, the choice of prior can make a relatively large difference. For example, in \cite{Bubeck2013}, a lower bound on the Bayesian regret of TS is established to be $\frac{1}{20}\sqrt{kn}$, along with a prior independent upper-bound of $14\sqrt{kn}$.

\begin{remark}
	While many frequentist bounds on the performance of TS exist, they generally focus on the asymptotic case, where the effects of the prior distribution ``wash out". In practice, performance can be highly suboptimal if the prior is misspecified and the number of observations is limited, as in a target tracking problem.
\end{remark}

Since the empirical performance of the TS strategy over a limited time horizon is highly dependent on the choice of prior distribution \cite{Bubeck2013,Dudik2021}, we discuss a procedure for sequentially learning a prior over time, when the radar is embedded in an environment of tasks coming from a common \emph{task distribution}. We will see that in many scenarios, meta-learning an effective prior for Thompson Sampling can provide major benefits in tracking performance. We now describe a meta-learning formulation in which uncertainty about the prior distribution is captured by a modeling a distribution over instance priors.

\section{Meta-Learning Formulation}
\begin{figure*}
	\centering
	\includegraphics[scale=0.5]{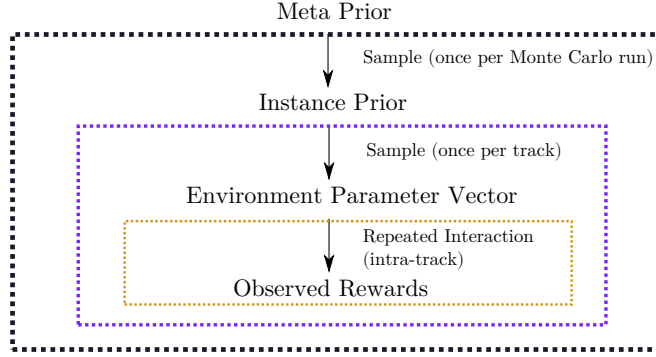}
	\caption{\textsc{Graphical flow} of the proposed meta-learning formulation. The instance prior $P_{\star}$ is sampled from $Q$ at the beginning of the simulation. The radar then engages in $m$ target tracking instances, where each target track consists of $n$ CPIs and is parameterized by $\theta_{s,\star} \sim P_{\star}$. After each target track, the radar updates its meta Posterior $Q_{s}$ using (\ref{eq:postupdateS}) and (\ref{eq:postupdate}) in order to obtain an estimate of $P_{\star}$.}
\end{figure*}

In this section, we describe a meta-learning procedure which aims to improve radar tracking performance using information gained over sequential interactions with $m$ stages of a Bayesian waveform selection problem. Each individual stage corresponds to a new instance of a $n$-step waveform-agile tracking problem on a FSTC described in Section \ref{se:problem}, where the radar has access to a catalog of $|\ncalW| = K$ waveforms. 

The radar ultimately wishes to learn the true parameter vector for stage $s$, denoted by $\theta_{\star} \in \mathbb{R}^{d}$. Given knowledge of $\theta_{\star}$, contexts can be directly mapped to average losses using the inner product relationship $\ell_{k} = \langle \theta, \varphi(o_{k},w_{k}) \rangle + \eta_{k}$ for stochastic linear bandits. As candidates, we consider the class of distributions $\mathcal{P} = \{P_{\theta} : \theta \in \Theta \}$, where $\Theta$ is a compact subset of $\mathbb{R}^{d}$. Just as in the conventional learning setting, the radar is equipped with a prior distribution $P(\theta)$ which expresses a subjective belief about the value of $\theta$ based on the current sequence of observations $\ncalH$.

At the beginning of each stage $s \in [m]$, an instance of the learning problem, specified by $\theta_{s,\star}$ is sampled from a \emph{task distribution} $P_{\star}$, which is fixed but unknown to the radar. Since $P_{\star}$ corresponds to the true distribution of tasks, each task being parameterized by $\theta_{s,\star}$, it can also be interpreted as the best possible choice of prior for a Bayesian learning algorithm. Thus, the meta-learning process consists of estimating the true prior $P_{\star}$ by sequentially interacting with problem instances that are assumed to be sampled i.i.d from a common distribution $Q$. 

The radar's uncertainty about the true value of $P_{\star}$ is reflected by assuming it is sampled from a fixed distribution over instance priors $P_{\star} \sim Q$. We refer to $Q$ as a \emph{meta-prior}\footnote{This is sometimes referred to as the \emph{hyper-prior} in the field of Bayesian statistics.}. The meta-learning environment is then characterized by the pair $(\ncalP,Q)$. We denote by $Q_{s}$ the \emph{meta-posterior}, which is the radar's current estimate of the instance prior $P_{\star}$ using information gained up to stage $s$. Good performance will occur when $D_{\texttt{KL}}(Q_{s}||P_{\star})$ is small, which corresponds to a low degree of uncertainty regarding the true prior $P_{\star}$. The meta-posterior is estimated sequentially using the standard Bayesian update rule
\begin{align}
	&Q_{s+1}(\hat{P}) \propto P(H_{s}|P_{\star} = \hat{P}) Q_{s}(\hat{P}) \nonumber \\ &= Q_{s}(\hat{P}) \int_{\theta} P(H_{s}|\theta_{s,\star} = \theta) P(\theta_{s,\star} = \theta | P_{\star} = \hat{P}) d\theta.
	\label{eq:metaUpdate}
\end{align}

The meta-learning formulation is an example of a hierarchical Bayes model.

\subsection{Meta-Thompson Sampling Contextual Bandit Algorithm}
To ensure that the meta-posterior update (\ref{eq:metaUpdate}) is computable in closed-form, and the existence of an efficient algorithm for online meta-learning, we consider a normal-normal conjugacy scenario, in which we study a Gaussian bandit with a Gaussian meta-prior. The Gaussian bandit model is commonly used in the literature on sequential decision processes over continuous parameter spaces. We assume a normal distribution over instance priors, expressed by $P(\theta) = \mathcal{N}(\mu,\sigma_{0}^{2}I_{d})$, where the noise level $\sigma_{0}^{2}$ is fixed. The meta-prior is then a distribution over instance prior means, given by $Q(\mu) = \mathcal{N}(\mathbf{0},\sigma_{q}^{2}I_{d})$. We assume the noise level $\sigma_{q}^{2}$ is fixed and known to the radar. The meta-learning process then maintains a meta-posterior $Q_{s}(\mu) = \ncalN(\hat{\mu}_{0,s},\hat{\Sigma}_{s})$. Due to the conjugacy properties of the normal distribution, the meta-posterior $Q_{s}$ can be simply updated in closed-form, even for a contextual bandit algorithm. Following the standard computations for Bayesian multi-task regression (reviewed in Appendix D of \cite{Kveton2021}), we see that the meta-posterior can be simply expressed by
\begin{equation*}
	Q_{s} \sim \ncalN(\mu_{s},\Lambda_{s}^{-1}),
\end{equation*}
where the parameters are updated each stage $s$ by
\begin{align}
	\label{eq:postupdateS}
	\Lambda_{s} &= \Lambda_{s-1} + X_{s}^{T}(\sigma^{2}I+X_{s}\Sigma X_{s}^{T})^{-1} X_{s}, \\
	\mu_{s}    &= \Lambda_{s}^{-1}(\Lambda_{s-1}\mu_{s-1}+ X_{s}^{T}(\sigma I) + X_{s} \Sigma X_{s}^{T})^{-1} L_{s}),
	\label{eq:postupdate}
\end{align}
where $\Sigma = \sigma_{0}^{2}I_{d}$, $X_{s} = [\varphi_{k=1},\varphi_{2},...,\varphi_{n}]$ is the vector of observed contexts over each time index $k = 1,2,...,n$ in stage $s$, and $L_{s} = [\ell_{k=1},\ell_{2},...,\ell_{n}]$ is similarly the vector of losses. The updates (\ref{eq:postupdateS}) and (\ref{eq:postupdate}) are easily calculated for low-dimension cases, such as the waveform selection problem examined here. For high-dimensional problems, the Woodbury matrix identity can be applied to speed up computations. A description of the meta-learning process can be seen in Algorithm \ref{algo:mts}.

We expect meta-learning to provide major benefits for cases where $\sigma_{q}^{2} \gg \sigma_{0}^{2}$, since uncertainty about the true prior $P_{\star}$ is significant. In cases where $\sigma_{q}^{2} \ll \sigma_{0}^{2}$, meta-learning is very close to the standard single-task learning problem, and we expect little benefit. Thus, the amount of relatedness between tasks is an important indicator of meta-learning performance. The issue of task similarity has been studied from an information-theoretic perspective in \cite{Jose2021}. Future work in this domain could focus on a rigorous justification of task similarity for radar tasks of practical interest, such as tracking under interference.

\begin{algorithm}[t]
		\setlength{\textfloatsep}{0pt}
	\label{algo:mts}
	\caption{Contextual Meta-TS for Waveform Agile Tracking}
	\SetAlgoLined
	\textbf{Input} meta-prior distribution $Q$, Loss function $\ell$\\
	Set $Q_{1} \leftarrow Q$ \\
	\For{\text{Each target track $s= \; 1,...,m$}}{
		\vspace{0.07cm}
		(1) Sample $P_{s} \sim Q_{s}$;\\ \vspace{.2cm}
		(2) Apply Thompson Sampling waveform selection policy (\ref{eq:tspolicy}) with prior $P_{s}$ to problem parameterized by $\theta_{s,\star} \sim P_{\star}$ for $n$ CPIs;\\ \vspace{0.2cm}
		(3) Update Meta-Posterior $Q_{s+1}$ according to the update rules (\ref{eq:postupdateS}) and (\ref{eq:postupdate}); \\
	}
\end{algorithm}

\section{Numerical Results}
\label{se:numerical}

\begin{figure}[t]
	\centering
	\includegraphics[scale=0.6]{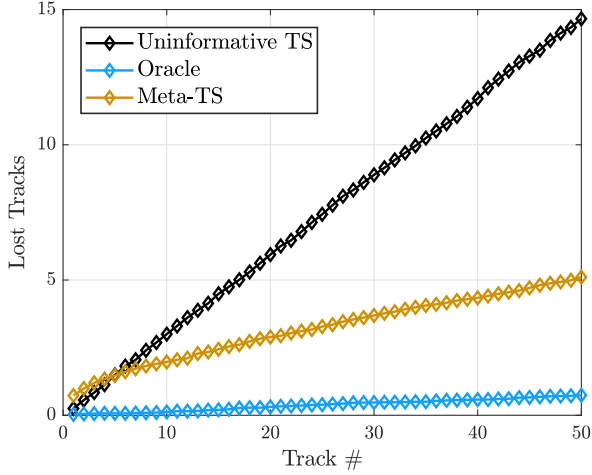}
	\caption{\textsc{Average number of cumulative lost tracks} for each learning algorithm. A lost track corresponds to 5 consecutive measurements where $\texttt{SINR} < 3 \texttt{dB}$. Results are Monte-Carlo averaged over 100 simulations.}
	\label{fig:lost}
\end{figure}

In this section, we examine the performance of the proposed meta-learning approach in a waveform-agile tracking problem that consists of $m = 50$ unique target tracks. Each target track corresponds to a new instance of a linear contextual bandit learning problem parameterized by $\theta_{s,\star}$ and consists of $n = 200$ radar CPIs, where each CPI corresponds to a decision round $k \in \nbbN_{+}$. The radar has access to a catalog of 5 waveforms. Once a waveform is selected, it is used throughout the CPI. At the end of each CPI, range-Doppler processing is performed using a matched filter and a measurement is extracted. The radar receives a loss given by
\begin{equation*}
\ell_{k} = -\texttt{SINR}_{\textrm{post},\textrm{dB}},
\end{equation*}
where the $\texttt{SINR}$ is estimated by averaging the energy in the estimated target bins of the range-Doppler map and averaging over the total energy in the map. We consider a case of $d = 2$, where the context features for each waveform consist of the average loss associated with transmitting that waveform as well as the worst-case loss associated with the waveform of interest, given an estimate of the target's position.

In Figure \ref{fig:lost}, we see the impact of meta-learning in terms of the number of tracks lost by each waveform selection algorithm. We consider a case of 5 consecutive measurements of $\texttt{SINR} < 3 \texttt{dB}$ to be a lost track. It is observed that the on average, the meta-TS algorithm loses almost the same number of tracks as when the true prior is known, and results in losing approximately four less tracks on average than applying Thompson sampling with an uninformative prior during each tracking instance. This can be attributed to a shortened exploration window early in the track, which is also when uncertainty about the target's position happens to be highest.

\begin{figure}[t]
	\centering
	\includegraphics[scale=0.55]{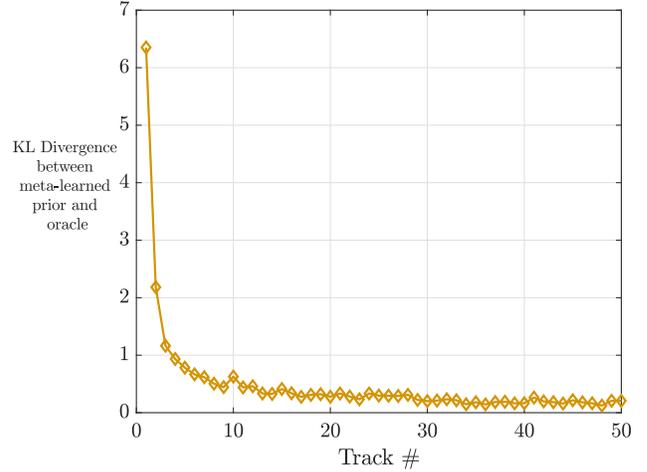}
	\caption{\textsc{KL divergence} between the meta-posterior $Q_{s}$ and true prior $P_{\star}$.}
	\label{fig:kl}
\end{figure}

Next, we would like to verify whether the meta-TS algorithm is learning the correct prior distribution. Figure \ref{fig:kl} shows the KL divergence between the meta-posterior $Q_{s}$ and the true prior $P_{\star}$. As this value tends toward zero, we expect the performance of the meta-learner to approach that of the oracle, which knows the true prior during each learning instance. We observe that after interacting with 10 tracks, the KL divergence becomes very small, indicating that $Q_{s}$ is approaching $P_{\star}$ and we can expect that meta-learning will perform similarly to the oracle.

\begin{figure}[t]
	\centering
	\includegraphics[scale=0.6]{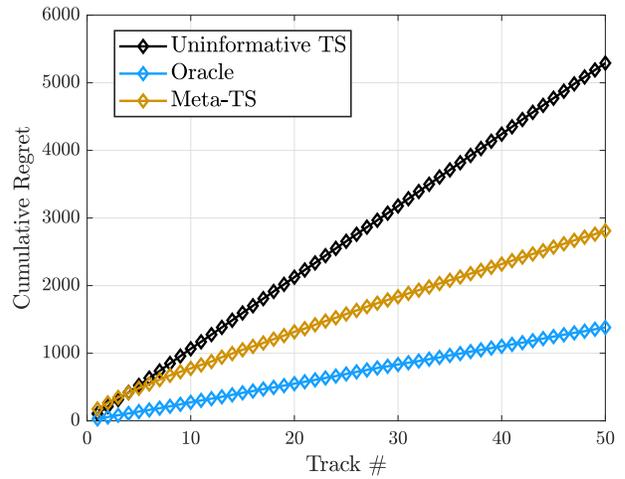}
	\caption{\textsc{Cumulative regret} for each waveform selection algorithm over $m = 50$ target tracks.}
	\label{fig:regret}
\end{figure}

Figure \ref{fig:regret} shows the cumulative regret incurred by each learning algorithm. As expected, we observe a notable gain from applying meta-learning in this setting, and the performance of the meta-leaner approaches that of TS with a known prior.

\begin{figure}[t]
  	\centering
  	\includegraphics[scale=0.6]{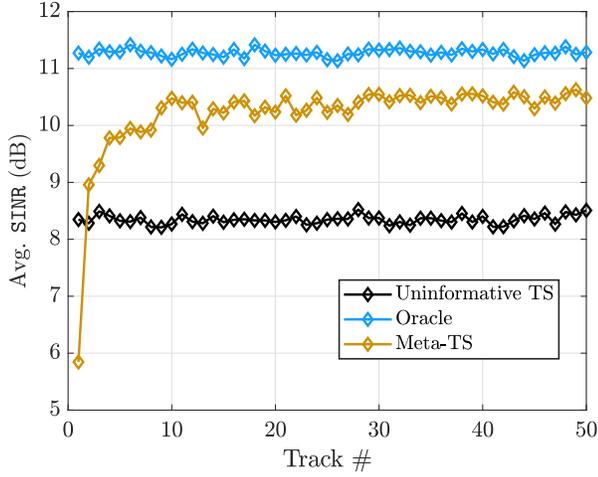}
  	\caption{\textsc{Average \texttt{SINR}} for each waveform selection algorithm over $m = 50$ target tracks.}
  	\label{fig:sir}
\end{figure}

\begin{figure}[t]
	\centering
	\includegraphics[scale=0.6]{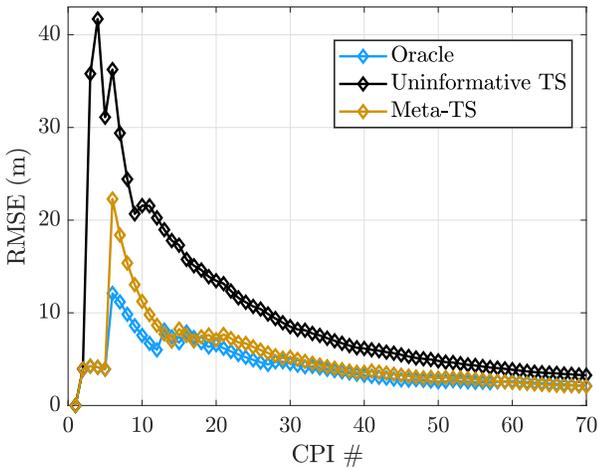}
	\caption{\textsc{RMSE} for each waveform selection algorithm over the final target track, $s = 50$.}
	\label{fig:rmse}
\end{figure} 

Figure \ref{fig:sir} shows the average $\texttt{SINR}$ of range-Doppler maps per track over the interval of $m = 50$ target tracks. We see that knowledge of the true prior is clearly beneficial in terms of average performance, and that meta-TS approaches the performance of the oracle, which applies TS with the best prior $P_{\star}$ during every track. However, the performance benefits in terms of average performance are not as substantial as in terms of worst-case metrics, such as lost tracks. This is because even when learning algorithms perform suboptimally, they often find solutions which result in good average performance. Thus, we expect meta-learning to provide the most substantial impact in terms of a reduced number of outages, which often occur early on in the learning process while the learner is exploring often.

In Figure \ref{fig:rmse} we examine the tracking RMSE of each algorithm over the final target target tracking instance, $s = 50$. We observe that the benefits associated with knowing the true prior occur early in the track when the tracking filter's uncertainty about the target's position is high. While all three learning approaches converge to a low RMSE, the meta-TS algorithm results in decreased error in the first few CPI's, in which the radar is most susceptible to losing the target track. We thus conclude that the most appreciable benefit meta-learning provides to tracking radar is a reduced risk of losing tracks due to the coupling high initial uncertainty about the target and the need for exploration.

\section{Conclusion and Future Directions}
This work has introduced a meta-learning approach for sample efficient online learning in a broad class of physical environments. We expect that meta-learning can provide appreciable benefits to cognitive radars, since these systems aim to adapt to an environment which is unknown \emph{a priori}, and an inductive bias can be found such that adaption can occur in all environments the radar may be exposed to in a timely manner. The meta-learning problem was framed using a hierarchical Bayes structure, and a linear contextual meta-Thompson Sampling algorithm was applied to learn an appropriate bias for a class of tracking objectives in a common environment. The proposed algorithm demonstrates a marked improvement over Thompson Sampling with an uninformative prior and performs nearly as well as the optimal case where the true prior is known.

While the current approach shows promise for many applications, it is not without shortcomings. To maintain tractability, we have assumed a normal-normal conjugacy model over the linear bandit problem. However, this structure does not seem to effect the generality of the approach, and we believe it should be appropriate for a realistic settings. However, for environments where the tracking tasks are nearly identical, we expect to see very little benefit from meta-learning relative to conventional learning approaches. 

Many forms of meta-learning have been proposed to handle online learning problems. There is significant potential for future work in comparing the utility of these various approaches for multi-task radar applications of practical value. An additional problem of interest is identifying practical deployments where meta-learning is expected to be useful. If the class of problems is too broad, the radar could potentially perform well with any choice of bias \cite{Jose2021}. For a single task, meta-learning is useless and utility maximization alone is sufficient. In other cases, learning may be extraneous as rule-based or random decision making may perform just as well. Future work could investigate how a meta-cognitive decision engine, such as \cite{Martone2020,Mishra2020} could be used to select the most appropriate level of learning abstraction for a given radar application. Finally, the computation of a Bayes optimal policy, which could be used to bound the performance of any decision agent may be tractable for certain scenarios and would be of theoretical interest. 
\bibliographystyle{IEEEtran}
\bibliography{metabib}

\end{document}